\documentclass[twocolumn,prb,showpacs]{revtex4}
\usepackage{graphicx}
\usepackage{epsfig}
\usepackage{amsmath}
\usepackage{amssymb,bm}
\usepackage{pstricks}

\usepackage{fancyhdr}
\usepackage{amsbsy}
\usepackage{amsmath}
\usepackage{subfigure}
\usepackage{epsfig}
\usepackage{amssymb}
\usepackage{pifont}
\usepackage{textcomp}
\usepackage{gensymb}
\usepackage{pifont}
\usepackage{enumerate}
\usepackage{color}
\usepackage{bm}
\usepackage{changes}

\begin{document}

\title{
Persistent spin helix in Rashba-Dresselhaus ferroelectric CsBiNb$_2$O$_7$
}

\author{Carmine Autieri$^{1,2}$, Paolo Barone$^{2}$, Jagoda S\l{}awi\'nska$^{3}$, and Silvia Picozzi$^2$}

\affiliation{$^1$International Research Centre MagTop, Institute of Physics, Polish Academy of Sciences, Aleja Lotnik\'{o}w 32/46, PL-02668 Warsaw, Poland}
\affiliation{$^2$ Consiglio Nazionale delle Ricerche, Istituto Superconduttori, Materiali Innovativi e Dispositivi (CNR-SPIN), c/o Universit{\`a} G. D'Annunzio, I-66100 Chieti, Italy}
\affiliation{$^3$Department of Physics, University of North Texas, Denton, TX 76203, USA}


\date{\today}

\begin{abstract}
Ferroelectric Rashba semiconductors (FERSC) are a novel class of multifunctional materials showing a giant Rashba spin splitting which can be reversed by switching the electric polarization. Although they are excellent candidates as channels in spin field effect transistors, the experimental research has been limited so far to semiconducting GeTe, in which ferroelectric switching is often prevented by heavy doping and/or large leakage currents. Here, we report that CsBiNb$_2$O$_7$, a layered perovskite of Dion-Jacobson type,  is a robust ferroelectric with sufficiently strong spin-orbit coupling and spin texture reversible by electric field. Moreover, we reveal that its topmost valence band's spin texture is quasi-independent from the momentum, as a result of the low symmetry of its ferroelectric phase. The peculiar spin polarization pattern in the momentum space may yield the so-called "persistent spin helix", a specific spin-wave mode which protects the spin from decoherence in diffusive transport regime, potentially ensuring  a very long spin lifetime in this material.
\end{abstract}

\maketitle

\section{Introduction}\label{sec:intro}
Novel materials with tunable spin properties hold promise to realize logic spintronics devices similar to spin field-effect transistors (spin-FET) proposed by Das and Datta.\cite{das} Recently discovered ferroelectric Rashba semiconductors (FERSC) emerge as excellent candidates for spin-FET channels, because of their unique coupling between spin and polar degrees of freedom enabling a purely electric non-volatile control of the electron's spin.\cite{silvia, fersc, liebmann} The Rashba spin splitting, due to inversion symmetry breaking with a single polar axis in these bulk crystals, can be controlled and permanently reversed by switching the sign of electric polarization, as  theoretically predicted and experimentally confirmed  in the FERSC prototype material GeTe.\cite{gete_nano} Moreover, being a bulk property, the Rashba spin texture in FERSC has been shown to be hardly affected when interfaced with ferromagnets,\cite{fe-gete, fert} which could be important when looking for robustness against the device configuration.

Unveiling peculiar properties of GeTe has stimulated several theoretical studies aiming at the discovery of novel FERSC. Several candidates have been proposed, ranging from similar chalcogenides (SnTe)\cite{plekhanov} to metal-organic halide perovskites, such as (FA)SnI$_3$,\cite{halides} hexagonal semiconductors (LiZnSb),\cite{narayan} and oxides (HfO$_2$, BiAlO$_3$, SrBiO$_3$).\cite{hfo2, luiz,sbo_prl2019} Most of them are robust ferroelectrics with sufficiently large band gaps preventing leakage currents and related problems with ferroelectric switching.\cite{fe-switching} These properties, together with large spin-orbit coupling (SOC), constitute essential requirements for realization of FERSC-based devices. On the other hand, strong SOC causes decoherence of electron spins in non-ballistic transport regime, making spin lifetimes of most FERSC too short for practical purposes.

Recent study on two-dimensional ferroelectric materials with in-plane electric polarization and out-of-plane spin polarization induced by SOC suggests that the problem of spin dephasing could be eliminated.\cite{hosik} In these materials, the symmetries of the crystal enforce the spin texture to be independent from the electron's momentum. Such spin configuration, called persistent spin texture (PST), was first proposed by Schliemann \textit{et al.}\cite{schliemann} for quantum wires with Dresselhaus and Rashba coefficients fine-tuned to have equal strengths. The corresponding spin wave mode emerging in the crystal, known as persistent spin helix (PSH),\cite{bernevig} protects the spins of electrons from dephasing, as they all precess at the same rate and in the same direction after scattering events. The concept of PSH has been recently extended to specific 3D bulk materials where the bulk crystal symmetries, rather than fine tuning of the SOC parameters, can substantially increase the spin lifetime.\cite{tsymbal_psh} Interestingly, a unidirectional spin-orbit field necessary to realize a PSH, has been recently predicted in a layered Aurivillius ferroelectric oxide, Bi$_2$WO$_6$, where it results from the combination of in-plane polarization and layering-related anisotropy of the electronic structure. \cite{bwo3_philippe}

Here, we reveal that another ferroelectric layered oxide, CsBiNb$_2$O$_7$, possesses a peculiar Rashba-Dresselahus spin texture, thus combining exceptional properties of FERSC with potentially long spin lifetime of carriers. CsBiNb$_{2}$O$_{7}$ is a layered perovskite of Dion-Jacobson type with a wide band gap, large ferroelectric polarization and SOC-derived spin splitting of the valence band of the order of 10 meV, which makes it a good FERSC candidate. Our density functional theory (DFT) calculations complemented by  $\bm k\cdot\bm p$ analysis have shown that the spin texture of the topmost valence band in the ferroelectric phase is close to the persistent spin-helix regime in large areas of the Brillouin zone. The model analysis suggests that in the ground state a small spin-momentum linear coupling term competes with third order coupling term, yielding a non-trivial spin pattern; on the other hand, for smaller ferroelectric displacements, a conventional Rashba spin texture is restored. Given that the Curie temperature for CsBiNb$_{2}$O$_{7}$ is expected to be much far the room temperature, the persistent spin helix emerges as a robust bulk property.

\begin{figure}[h]
\includegraphics[width=\columnwidth]{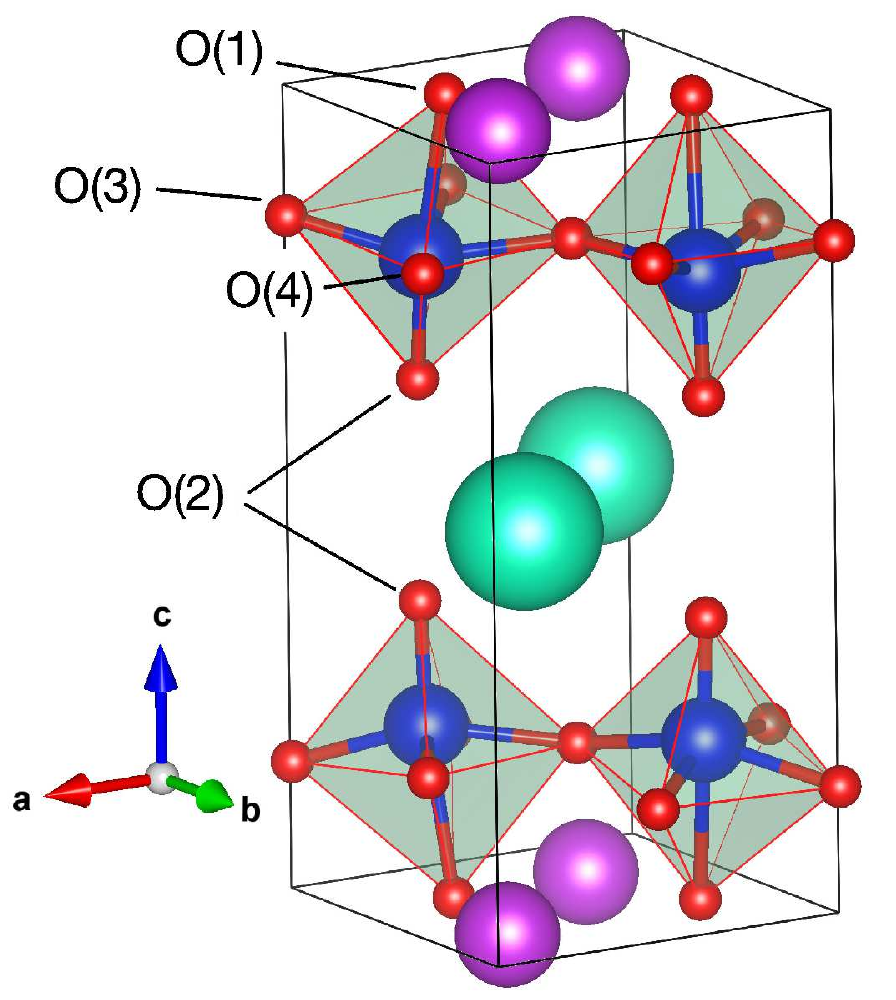}
\caption{Crystal structure of the CsBiNb$_2$O$_7$. The NbO$_6$ octahedra are shown in light green. They are separated along the $c$-axis by Cs atoms displayed as large green spheres. The Bi atoms in Bi-O planes are marked as violet spheres. O(1)-O(4) denote inequivalent oxygen atoms (small red spheres).}\label{fig:CS}
\end{figure}

The paper is organized as follows. In Section \ref{sec:computational}, we describe details of the DFT calculations. In Section \ref{sec:results_electronic}, we discuss the electronic properties and the ferroelectricity of CsBiNb$_2$O$_7$, while Sections \ref{sec:results_dftspins} and \ref{sec:results_kpspins} report the spin textures estimated from DFT and analyzed in the framework of $\bm k\cdot\bm p$ theory. Section \ref{conclusions} summarizes the conclusions and perspectives of further research.

\section{Computational details}\label{sec:computational}

We have performed first-principles DFT
calculations employing the VASP package \cite{vasp1,vasp2} based
on plane wave basis set and projector augmented wave method.\cite{paw1,paw2}
We have used a plane-wave energy cut-off of 480~eV and
generalized gradient approximation of Perdew-Burke-Ernzerhof as the exchange-correlation functional.\cite{pbe}
We have set the experimental internal positions and lattice constants equal to $a$=5.49528 {\AA}, $b$=5.42251 {\AA} and $c$=11.37663 {\AA}.\cite{positions} For Brillouin zone integrations, a $8 \times 8 \times 4$ k-points grid has been used. An additional Coulomb repulsion within a DFT+U approach  on the Nb atoms has been neglected, because it was estimated to 1-3 eV in $4d$ perovskite oxides,\cite{Autieri4dU1,Autieri4dU2} which is an intermediate value that increases the band gap, but only slightly affects the band structure with empty $d$-shell. We also note that the effect of the Coulomb repulsion would be even less relevant on the Bi-O states in the valence band as they weakly hybridize with the Nb states. The electric polarization has been calculated using the Berry phase method.\cite{polarization} The spin-orbit coupling (SOC) has been included in all the calculations.

\section{Results}

\begin{figure}[h!]
\includegraphics[height=\columnwidth, angle=270]{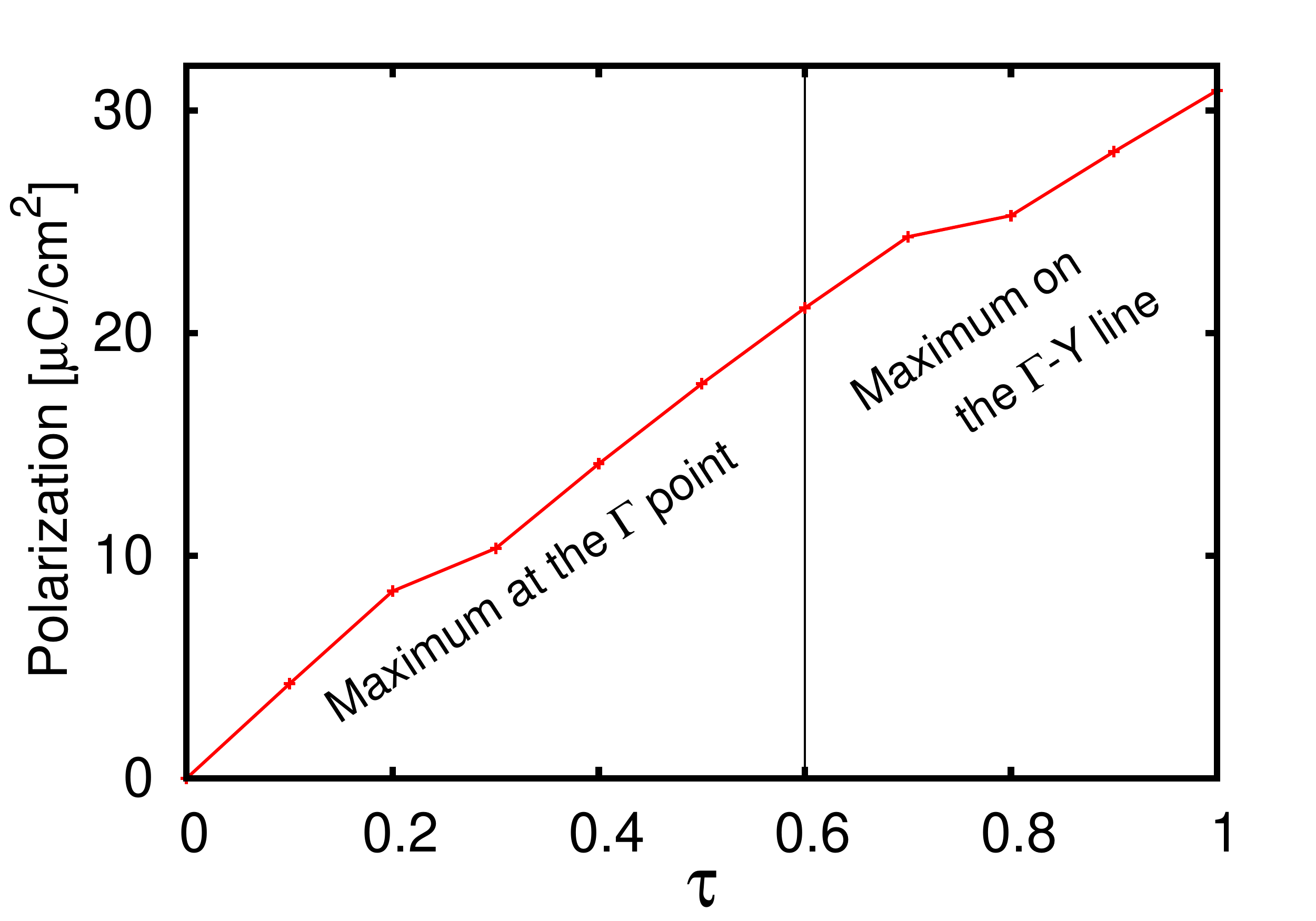}
\caption{Total polarization along the ferroelectric path. 
The vertical line at $\tau=0.6$ separates two regions with qualitatively different dispersion of the valence band.
}\label{fig:tau}
\end{figure}

\begin{figure*}[ht!]
\includegraphics[width=\textwidth]{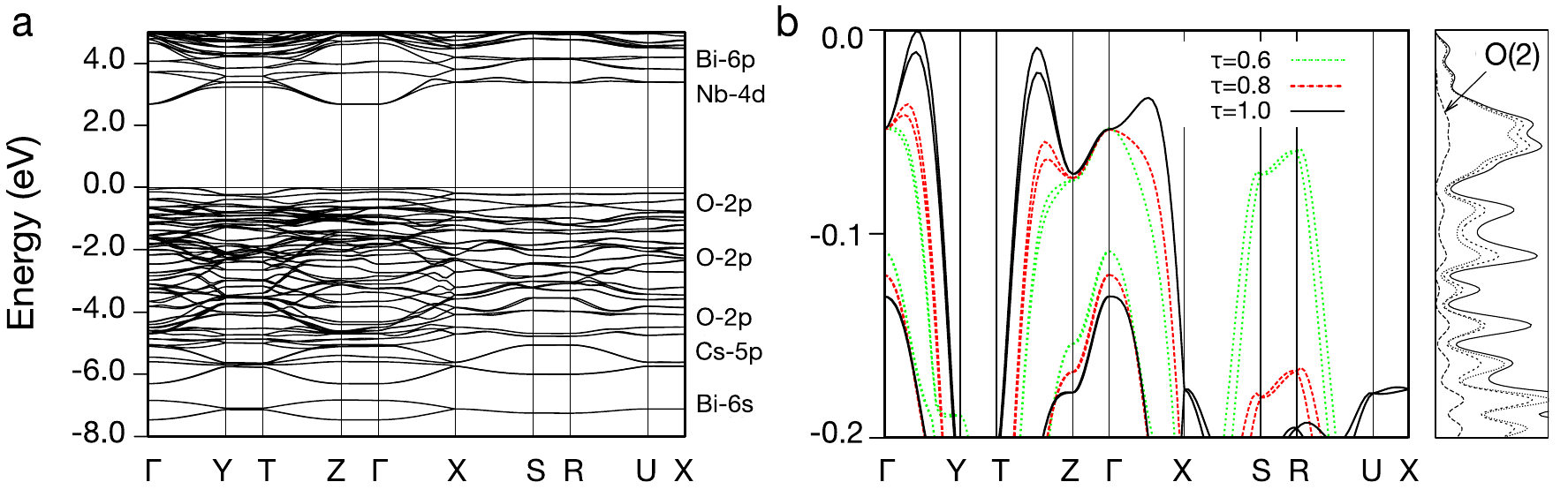}
\caption{(a) Band structure of the CsBiNb$_2$O$_7$. The Fermi level is set to zero. (b) Band structure of the topmost valence band calculated for $\tau$=0.6 (dotted green line), 0.8 (dashed red line) and 1.0 (solid black line).
The projected DOS of the oxygen atoms is shown as side panel. The solid, dashed, short dashed and dotted black lines represent the DOS of the O(1), O(2), O(3) and O(4) atoms, respectively.}
\label{fig:bands}
\end{figure*}

\subsection{Electronic properties: band structure and polarization}\label{sec:results_electronic}

Figure \ref{fig:CS} shows the crystal structure of CsBiNb$_2$O$_7$. Previous studies revealed that it is an $n=2$ Dion-Jacobson ferroelectric belonging to the space group $P2_1am$ (no. 26) with the in-plane electric polarization parallel to the $a$ axis.\cite{fennie} The contributions of different layers to the ferroelectricity evaluated for this class of materials showed that the origin of polarization is mainly related  to the BiO layers.\cite{benedek} Our calculated polarization value of 30.9 $\mu$C/cm$^2$ (see Fig. \ref{fig:tau}) obtained using the experimental atomic positions of CsBiNb$_2$O$_7$ is approximatively 20\% smaller than the values found in the literature for this and a similar compound RbBiNb$_2$O$_7$, \cite{fennie, band_structure,benedek} which could be attributed to different lattice constants and exchange functionals.




First, we focus on the electronic dispersion within the crystallographic planes orthogonal to the polar axis $a$, where we expect the largest SOC-derived spin-splitting effects. 
Nominally, the Nb atoms are in a 4d$^0$ configuration while the Cs and Bi atoms are in a 6$s^0$ and 6p$^0$ configurations, respectively. Finally, all the oxygen atoms are in a 2p$^6$ configuration. Figure \ref{fig:bands} (a) shows the band structure,  plotted along the k-paths lying in two high-symmetry planes orthogonal to the polar axis $a$. We found a band gap of 2.7 eV, in qualitative agreement with the band structure of RbBiNb$_2$O$_7$,\cite{band_structure} another compound of the same class. In the polar phase, the band gap is indirect with the minimum of the conduction band at $\Gamma$ and the maximum of the valence band along the $\Gamma$-Y line. We  found that the dispersion along the z-axis perpendicular to the layers is reduced by almost an order of magnitude with respect to the in-plane dispersion, similar to other $4d$ layered perovskites.\cite{Autieri2012,Autieri2014}



The orbital composition of the bands is displayed on the right-hand side of Fig. \ref{fig:bands}(a). We  found that between -7.5 eV and -5.5 eV, the electronic structure is mostly of Bi-$6s$ and Cs-$5p$ character, respectively. From -5.5 eV up to the Fermi level, the dispersion is dominated by the oxygen states. We note that there are 4 inequivalent oxygen atoms in the unit cell, as shown in Fig. \ref{fig:CS}. The O(1) and O(2) atoms are the apical oxygens of the NbO$_6$ octahedra;  they are qualitatively different, since O(1) belongs to the Bi plane and O(2) does not. The O(3) and O(4) atoms are the planar oxygens of the NbO$_6$ octahedra and they are both first neighbors of the Bi atoms. The density of states just below the Fermi level is contributed by the atoms O(1), O(3) and O(4) strongly hybridizing with  Bi states. The contribution from the atoms O(2) is not relevant, as shown in the DOS of Fig. \ref{fig:bands}(b). Finally, the bottom of the conduction band is composed by Nb-$4d$ bands, while the higher conduction bands display the Bi-$6p$ character.


As a next step, we have investigated in more detail the topmost valence band, which is the most interesting part of the electronic structure. Figure \ref{fig:bands} (b) presents the dispersion energy region closest to the Fermi level (black lines); we can observe a non-negligible spin-splitting close to the $\Gamma$ point, namely along the lines $\Gamma$-Z and $\Gamma$-Y corresponding to the direction perpendicular to the polar axis in the reciprocal space. Since the system is layered, the splitting is much larger within the basal plane along the $\Gamma$-Y line. The spin-orbit splitting at the maximum of the valence band achieves 10 meV. In addition, we have analyzed the evolution of the valence band versus the ferroelectric displacement, defined as $\tau$=0 for the paraelectric structure and $\tau$=1 for the experimental ferroelectric phase (see Fig. \ref{fig:bands}(b)). It is clear that the position of the valence band maximum (VBM) strongly depends on the ferroelectric distortion. A decrease in $\tau$ reduces the spin-orbit splitting and moves the position of the VBM towards $\Gamma$.
At the same time, the topmost valence bands in the plane $k_x=\pi/a$ are shifted to lower energies. We have found that the change of behavior occurs at $\tau\sim$0.6; below this value, the position of the VBM is at the $\Gamma$ point, while for $\tau>$0.6 it gets shifted along the $\Gamma$-Y line. As a consequence, the gap changes from direct to indirect as a function of the ferroelectric displacement. The Bi- and O-like valence bands are strongly influenced by SOC, while the lowest Nb conduction bands remain almost unaffected.


\subsection{From the Rashba spin texture to the persistent spin helix}\label{sec:results_dftspins}

\begin{figure}[h]
\includegraphics[width=\columnwidth]{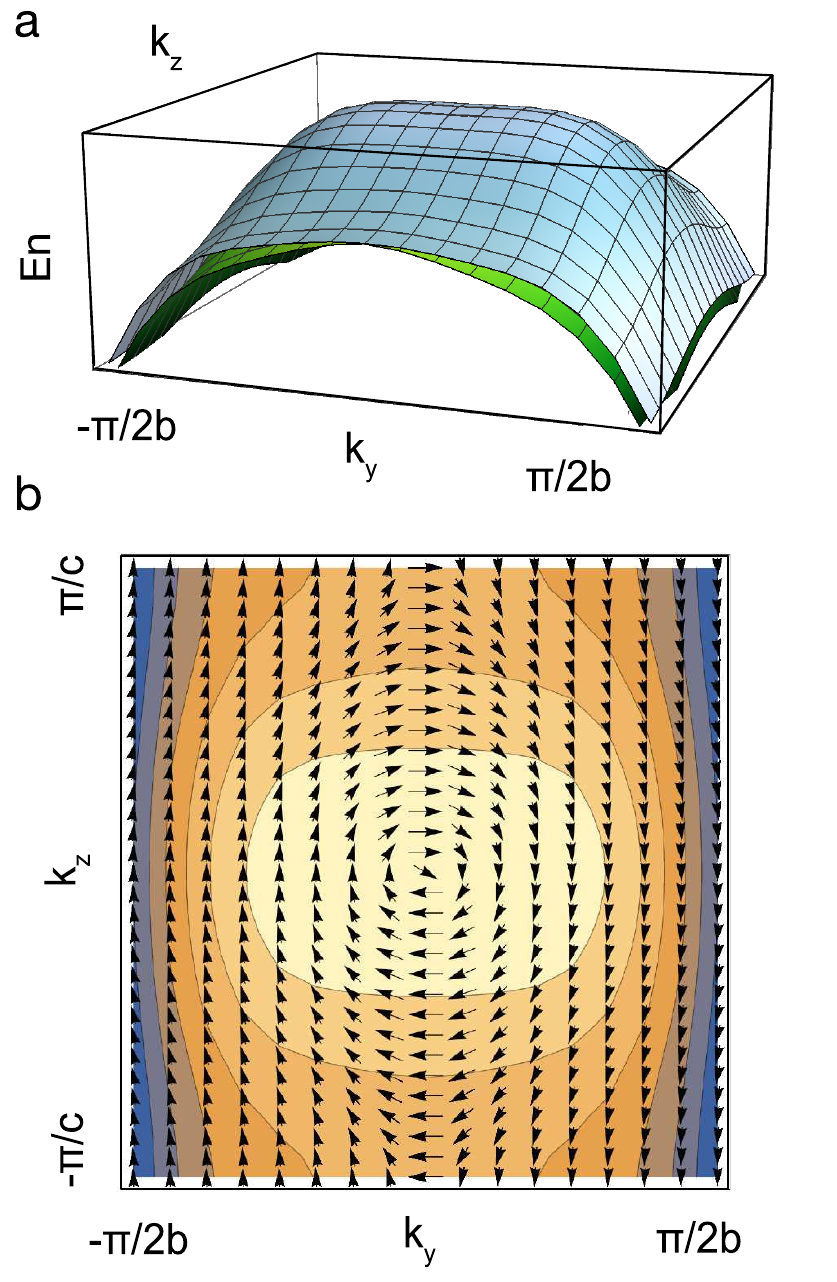}
\caption{(a) Spin-split energy bands induced by the spin-momentum coupling at $\tau$=0.6 calculated from first principles. (b) Corresponding spin texture of the valence band at $\tau$=0.6. Arrows denote in-plane spins whereas the color scale shows the energy difference from the Fermi level. The light yellow denotes the regions close to the Fermi level, while dark blue denotes the regions far from the Fermi level.}\label{fig:Ektau06}
\end{figure}

\begin{figure}[h]
\includegraphics[width=\columnwidth]{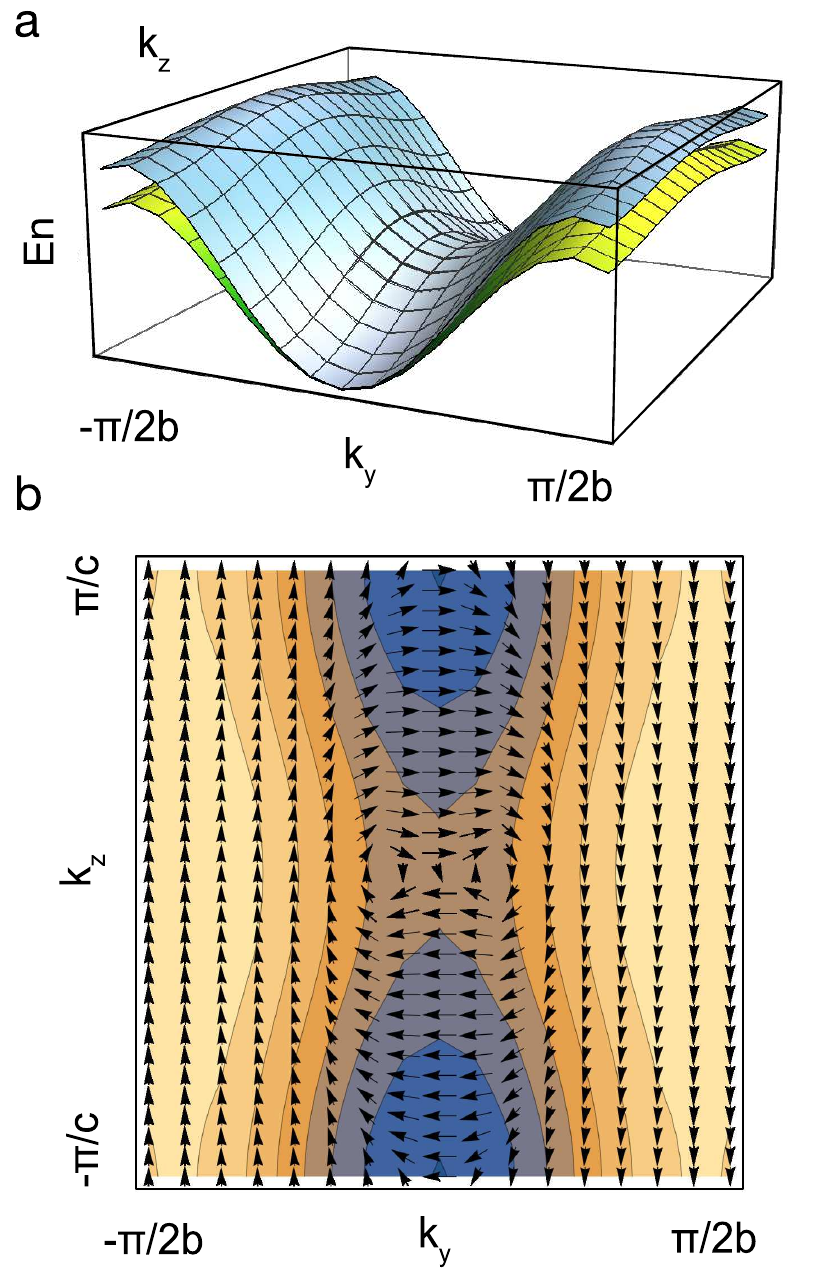}
\caption{(a) Spin-split energy bands induced by the spin-momentum coupling for $\tau$=1.0 calculated from first-principles. (b)
Corresponding spin texture of the valence band at $\tau$=1.0. Arrows denote in-plane spins whereas the color scale shows the energy difference from the Fermi level. The light yellow denotes the regions close to the Fermi level, while dark blue denotes the regions far from the Fermi level.}\label{fig:Ektau10}
\end{figure}

We have further investigated the evolution of the energy dispersion and the corresponding spin texture of the topmost valence band as a function of the ferroelectric displacement $\tau$. We have focused on the most interesting area of the Brillouin zone at k$_x$=0. Figure \ref{fig:Ektau06}(a) shows the valence band for $\tau$=0.6 in the k$_y$ range between -$\frac{\pi}{2b}$ and $\frac{\pi}{2b}$ and in the k$_z$ range between -$\frac{\pi}{c}$ and $\frac{\pi}{c}$. The dispersion reveals a negative effective mass both along the $k_y$ and $k_z$ axis, as well as a strong dependence on k$_y$ and a weak dependence on k$_z$
(most likely due to the layered structure of CsBiNb$_2$O$_7$). Although this anisotropy is reflected also in the spin texture [cfr Fig. \ref{fig:Ektau06}(b)], the latter closely resembles the conventional Rashba-like one. As can be observed in Fig. \ref{fig:Ektau10}(a), at $\tau$=1.0 the dispersion is dramatically different. The effective mass is still negative along the $k_z$, but it becomes positive along the $k_y$ direction. Even a more striking change can be noticed in the spin texture reported in Fig. \ref{fig:Ektau10}(b), especially at small momenta around $\Gamma$, where we have found a complex pattern of spin polarization, different either from Rashba- and Dresselhaus-like spin textures. The most important feature, however, is the spin fixed along the z-direction in a wide region around  $k_y\approx\pm\frac{\pi}{2b}$, which gives rise to a persistent spin texture pattern.

Before analyzing in detail the mechanisms driving such spin polarizations, let us briefly compare our results with known 3D crystals in which PST is enforced by the symmetries. For example, in prototype BiInO$_3$ and recently proposed Bi$_2$WO$_6$, the band splittings are substantially larger, 260 and 70 meV, respectively.\cite{tsymbal_psh, bwo3_philippe} This could make them more suitable for room temperature functionalities than CsBiNb$_2$O$_7$ with spin splitting of only 10 meV. However, the new and intriguing property of the latter is the PSH residing in the valence band, which opens a perspective for its direct observation using spin- and angle-resolved photoemission spectroscopy (ARPES) without the need of any additional doping. Furthermore, despite the SOC in CsBiNb$_2$O$_7$ not being perfectly unidirectional over the entire BZ, we note that the previously studied ferroelectric oxides also reveal deviations from the ideal PST. Thus, we are convinced that the pattern shown in Fig. \ref{fig:Ektau10}(b) can still result in a reduction of spin decoherence in the diffusive transport regime in this material.


\subsection{$\bm k \cdot \bm p$ analysis of valence-band spin polarizations}\label{sec:results_kpspins}

In order to understand the SOC-induced spin-splitting and the non-trivial spin textures of CsBiNb$_2$O$_7$, we derive
 a minimal $\bm k\cdot\bm p$ model for $J=1/2$ states, enforcing the specific symmetry properties of the system.
The point group of the $\Gamma$ point is $C_{2v}$, consisting in a two-fold rotations around the polar axis (chosen parallel to $x$, consistently with the $P2_1am$ setting used in DFT calculations) and two mirror operations about two  planes containing the polar axis. We notice that coexisting Rashba and Dresselhaus couplings are generally allowed within this point-group symmetry\cite{halides}, which may lead to PSH once Rashba and Dresselhaus coupling constants compensate exactly\cite{schliemann}.
The calculated electronic structure further suggests that the linear spin-momentum coupling is quite small, the spin-splitting being dominated by the third-order coupling (see, e.g., Fig. \ref{fig:bands}(b)).
%
The minimal model $H_{RD}$ with Rashba and Dresselhaus couplings up to third-order in momentum, which includes all symmetry-allowed terms, reads as:

\begin{eqnarray}\label{eq:ham}
H_{RD}&=&E_0+\alpha(k)\,k_y\sigma_z +\beta(k)\,k_z\sigma_y
\end{eqnarray}
\noindent
where $k_y,k_z$ are the cartesian components of crystal momentum in the plane perpendicular to the polar axis $a$ and $\bm \sigma$ are Pauli matrices describing spin degrees of freedom in the same reference system.
The free-electron (parabolic) contribution is $E_0=\hbar^2k_y^2/2m^*_y-\hbar^2k_z^2/2m^*_z$, where we assumed opposite effective masses $m^*_y$, $m^*_z$ along the orthogonal directions $k_y$ and $k_z$,  consistently with the DFT calculations.

The $k-$dependent spin-orbit coupling constants in the chosen reference frame are
\begin{eqnarray}
\alpha(k)&=&\alpha^{(1)}+\alpha^{(3)} k^2 +\gamma_\alpha\,(k_y^2-k_z^2)\\
\beta(k)&=&\beta^{(1)}+\beta^{(3)} k^2+\gamma_\beta\,(k_y^2-k_z^2),
\end{eqnarray}
where $\alpha^{(3)}$, $\beta^{(3)}$ are $k^2$ renormalization terms of the linear spin-orbit coupling constants, while $\gamma_\alpha, \gamma_\beta$ account for the $k$-cubic anisotropic interactions.

\begin{figure*}[ht]
\includegraphics[width=0.48\textwidth]{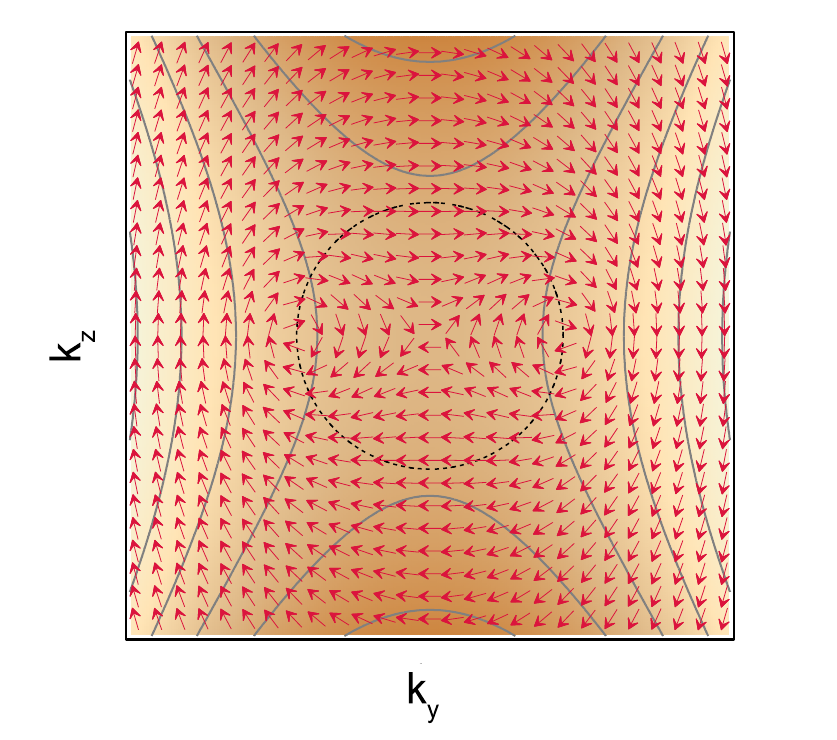}
\includegraphics[width=0.48\textwidth]{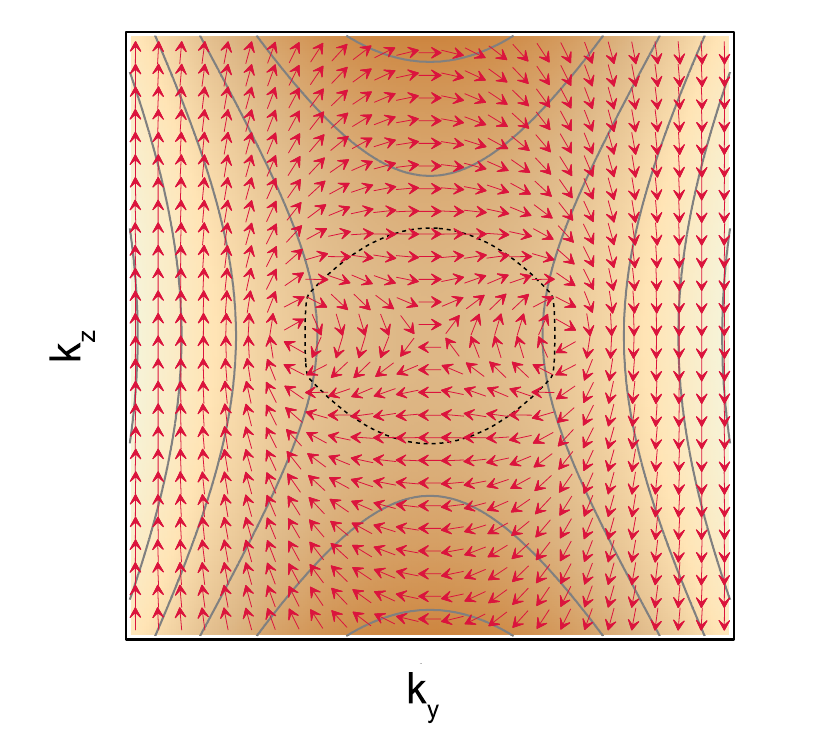}
\caption{Spin-polarization pattern evaluated from the $\bm k$ $\cdot$ $\bm p$ model Eq. (\ref{eq:ham}) with different choices of the model parameters compatible with the results of the fitting procedure given in Table \ref{TabFitting}. (Left panel) Spin texture for $\gamma_\alpha=\gamma_\beta=0$ and $\alpha^{(3)}=21.6$ eV{\AA}$^3$, $\beta^{(3)}$=-2.52 eV{\AA}$^3$. (Right panel) Spin texture for $\gamma/2=\gamma_\alpha=-\gamma_\beta=-5.2$ eV{\AA}$^3$, $\alpha^{(3)}=26.8$ eV{\AA}$^3$ and $\beta^{(3)}$=2.68 eV{\AA}$^3$, corresponding to an anisotropy factor $f=0.1$.
Since the $\bm k \cdot \bm p$ model is  by construction accurate only around $\Gamma$, we plot the pattern in a small range around the Brillouin zone center.
The dashed black loop around $\Gamma$ highlights the $k$ value where $\alpha(k)$ changes sign; the spin-polarization pattern inside (outside) the loop displays a Dresselhaus-like (Rashba-like) character.}
\label{fig:spin}
\end{figure*}

The explicit expressions of spin expectation values for model (\ref{eq:ham}) can be easily derived, reading:
\begin{eqnarray}
\langle \sigma_y^\pm \rangle &=&\pm \frac{\beta(k) k_z}{\sqrt{\alpha(k)^2\,k_y^2+\beta(k)^2\,k_z^2}} \nonumber\\
\langle \sigma_z^\pm \rangle &=&\pm \frac{\alpha(k) k_y}{\sqrt{\alpha(k)^2\,k_y^2+\beta(k)^2\,k_z^2}}, \label{eq:spinpol}
\end{eqnarray}
which show the dependence of the spin directions onto the sign of the $k$-dependent coupling constants $\alpha(k),\beta(k)$. The pure linear Rashba (Dresselhaus) limit is recovered when $\alpha^{(1)}=-\beta^{(1)}$ ($+\beta^{(1)}$) and $\alpha^{(3)}=\beta^{(3)}=0$, while the cubic Dresselhaus term is obtained for $\gamma_\alpha=-\gamma_\beta$. A Rashba-like (Dresselhaus-like) spin texture is also expected when the $k-$dependent coupling constants $\alpha(k)$, $\beta(k)$ display the same relative signs as above, the spin-polarization pattern being possibly modulated by the anisotropy of the coupling terms.
On the other hand, the persistent spin helix would be realized if $\alpha(k)$ or $\beta(k)$ vanishes, leaving a unique spin quantization axis ($\hat{y}$ or $\hat{z}$, respectively).



Aiming at a quantitative analysis, we fitted the model parameters from
the sum and the difference between the two topmost valence bands 
along the $\Gamma$-Y and $\Gamma$-Z lines. The effective masses are then obtained by fitting the sum of the two topmost valence bands,
while from the difference we get the odd terms  $\alpha^{(1)}$, $\alpha^{(3)}+\gamma_\alpha$,
$\beta^{(1)}$ and $\beta^{(3)}-\gamma_\beta$
related to the spin-orbit parameters.
Results are reported in Table \ref{TabFitting}. 
Unfortunately, we were not able to determine accurately the single values of  $\alpha^{(3)}$, $\beta^{(3)}$, $\gamma_{\alpha}$ and $\gamma_{\beta}$, which would have required to fit the bands away from the high-symmetry $\Gamma$-Y and $\Gamma$-Z lines, introducing additional numerical uncertainty.
Nevertheless, the linear spin-momentum coupling constants are two to three order of magnitude smaller than the cubic ones. On the other hand, assuming for simplicity that $\gamma_\alpha=\gamma_\beta=0$, the coupling constant $\alpha^{(3)}$ is one order of magnitude larger than $\beta^{(3)}$, explaining the partial compensation of Rashba and Dresselhaus effects and confirming the almost unidirectional spin-orbit field at large momenta along the $\Gamma$-Y line.
Finally, the opposite signs of the small linear $\alpha^{(1)}$ and cubic $\alpha^{(3)}$ coupling constants can explain the observed non-trivial spin-polarization patterns at small momenta. In fact, as long as $\vert\alpha^{(1)}\vert >\vert\alpha^{(3)}\,k^2\vert$  - i.e., very close to the $\Gamma$ point - the two $k$-dependent coupling constants $\alpha(k)$ and $\beta(k)$ display the same sign, being both negative, thus supporting a Dresselhaus-like spin-texture; the Rashba-like features emerge again at larger momenta, where the sign of $\alpha(k)$, dominated by the cubic $\alpha^{(3)}$ coupling constant, is opposite to that of $\beta(k)$. As a consequence, the spin-polarization pattern shown in Fig. \ref{fig:spin}(left panel) is realized, which already agrees qualitatively well with the DFT results shown in Fig. \ref{fig:Ektau10}(b). We notice that such peculiar spin-texture could not be reproduced by neglecting the $k^2$ renormalization terms, i.e., for $\alpha^{(3)}=\beta^{(3)}=0$ and assuming the k-cubic spin splitting as arising solely from the SOC parametrized by $\gamma_\alpha,\gamma_\beta$. In order to get some insight on the role of such $k$-cubic anisotropic term, and having assessed the primary role of the anisotropy of $\alpha^{(3)}$ and  $\beta^{(3)}$, we checked how the spin-polarization pattern is affected by including a cubic Dresselhaus term (parametrized by $\gamma/2=\gamma_\alpha=-\gamma_\beta$) and assuming $\beta^{(3)}=f\alpha^{(3)}$, where $f$ is an arbitrary independent parameter quantifying the anisotropy of $\alpha^{(3)}$ and $\beta^{(3)}$ $k^2$ renormalization terms. We found that the qualitative agreement between the effective model and the DFT results slightly improves for $f$ between -0.1 and 0.1, as shown in Fig. \ref{fig:spin}(right panel), suggesting that an almost unidirectional spin-orbit field at large momenta might be further stabilized by a cubic Dresselhaus-like spin-momentum coupling\cite{PhysRevB.86.081306}.

\begin{table}[!ht]
\caption{Ab-initio values of the kinetic and spin-orbit parameters, evaluated from a fit close to the $\Gamma$ point.
The quantities $\frac{\hbar^2}{2m_y}$  and $\frac{\hbar^2}{2m_z}$  are in eV{\AA}$^2$.
The units are eV{\AA} and eV{\AA}$^3$, respectively, for the first and third order terms in k.}
\begin{center}
\renewcommand{\arraystretch}{1.4}
\begin{tabular}{|c|c|c|c|c|c|}
\hline
     \multicolumn{2}{|c}{kinetic parameters} & \multicolumn{4}{|c|}{SOC parameters } \\
\hline
$\frac{\hbar^2}{2m_y}$ & $\frac{\hbar^2}{2m_z}$ & $\alpha^{(1)}$ & $\beta^{(1)}$ & $\alpha^{(3)}+\gamma_{\alpha}$ & $\beta^{(3)}-\gamma_{\beta}$ \\
\hline
\parbox{1.2cm}{15.9}  &  \parbox{1.2cm}{8.59}  &  \parbox{1.2cm}{-0.0102} & \parbox{1.2cm}{-0.0145} & \parbox{1.2cm}{21.6} & \parbox{1.2cm}{-2.52} \\
\hline
\end{tabular}
\end{center}
\label{TabFitting}
\end{table}

\section{Conclusions}\label{conclusions}
In summary, we unveiled that the Dion-Jacobson layered perovskite CsBiNb$_{2}$O$_{7}$ possesses the unidirectional spin texture independent from the momentum over large areas of the Brillouin zone, which gives rise to the persistent spin helix and potentially long spin lifetimes. We emphasize that this peculiar spin wave mode emerges as an intrinsic property of the bulk phase, thus does not rely on fine tuning of Rashba and Dresselahus parameters which was a strong requirement to realize the PSH in semiconductor quantum wells explored in last years.\cite{Koralek2009,PhysRevB.86.081306,walser,ap_express,passmann,altmann}

In particular, we have shown that the crystal symmetry (space group $P2_1am$) and the point group of the $\Gamma$ point ($C_{2v}$) around which the spin-splitting occurs, perfectly match the requirements for the PSH design, as it allows for coexisting Rashba and Dresselhaus couplings. Our DFT calculations reveal that these SOC parameters compensate well and the spin texture can be considered quasi-independent from the momentum. 

Finally, in contrast to similar ferroelectric oxides the spin-splitting is found in the valence band. Despite being rather small (10 meV) and related to higher-order spin-momentum coupling, it yields PSH which could be directly measured via ARPES, provided a sufficiently accurate resolution is reached. Moreover, as CsBiNb$_2$O$_7$ is a robust ferroelectric and an example of FERSC, a ferroelectric switching of PSH could be observed as well. We hope our results will be useful for realization of novel spintronics devices based on FERSC and will stimulate further search of similar materials with Rashba-Dresselhaus spin textures that could yield a persistent spin helix and, ideally, infinite spin lifetimes.

\section*{Acknowledgments}
We thank Luiz Gustavo Davanse da Silveira and Sang-Wook Cheong for useful discussions. The work is supported by the Foundation for Polish Science through the IRA Programme co-financed by EU within SG OP.
C.A. acknowledges the CINECA award under the ISCRA initiative IsC54 "CAMEO" and IsC69 "MAINTOP" Grant, for the availability of high performance computing resources and support.

\bibliography{Rashba_CsBiNb2O7}




\end{document}